\documentclass[preprint, showpacs,preprintnumbers,amsmath,amssymb]{revtex4-1}

\usepackage{xspace}
\usepackage{epsfig}
\usepackage{amsfonts}
\usepackage{amsmath}
\usepackage{graphicx}
\usepackage{url}
\usepackage{color}
\usepackage{afterpage}

\begin{document}

\title{A tunnel-field-effect spin-filter from two-dimensional anti-ferromagnetic stanene}

\author{E. G. Marin}
\thanks{These authors contributed equally to this work.}
\author{D. Marian} 
\thanks{These authors contributed equally to this work.}
\author{G. Iannaccone} 
\author{G. Fiori}
\email{email: g.fiori@iet.unipi.it}

\affiliation{Dipartimento di Ingegneria dell'Informazione, Universit\`{a} di Pisa, Pisa, 56122, Italy.}

\begin{abstract} We propose a device concept, based on monolayer stanene, able to provide 
highly polarized spin currents (up to a $98\%$) with voltage-controlled spin polarization
operating at room temperature and with small operating voltage ($0.3$~V). 
The concept exploits the presence of spin-polarized edge states in a stanene nanoribbon. 
The spin polarization of the total current can be modulated by a differential tuning of 
the transmission properties, and of the occupation of edge states of different spin, 
via the application of an in-plane electric field. We demonstrate device operation using
ab-initio and quantum transport simulations.
\end{abstract}
\vskip-5ex

\maketitle

\section*{Introduction}

Spintronics, the application of the electron spin degree of freedom to 
information technology ~\cite{Wolf01,Awschalom07,Felser07}, has experimented an 
impressive progress since the discovery of the giant magneto-resistive effect in 
late eighties~\cite{Baibich88}{, that was followed and improved, a few years later, by the tunneling magnetoresistance effect \cite{Modera1995}.  Nowadays, spintronics constitutes a intense research area embracing from physics to computer science \cite{Xu2016,Zutic2004}, and involving exciting and promising fields as topological insulators \cite{Qi2011} and quantum computing \cite{Leuenberger2001}.}

{Indeed, although many issues are still at their infancy including 
efficient injection, transport, and control of spin currents \cite{Avci17,Miron11}, 
enormous progresses have been made, and some notable devices have been proposed and demonstrated, such as, e.g.  spin torque memories,  \cite{Hosomi05,Liu12,Prenat2015} spin valves in hard drives read heads \cite{Tsang1994}, galvanic isolators \cite{Hermann1997}, magnetic memories based on tunnel junctions \cite{Gallagher2006}, spin transistors \cite{Datta1990} and logic devices built from diluted magnetic semiconductors \cite{Ohno2000}.}

Further progresses are expected to be enabled by the investigation of promising combinations of materials and structures
\cite{Gao13,Ohno16}, including  ferromagnetics stacks \cite{Jedema01}, half-metals \cite{deGroot83,Son06}, or spin-gapless semiconductors
\cite{Wang08,Ouardi13}. More recently, the raise of graphene~\cite{Novoselov04} opened new possibilities in the field \cite{Roche15,Han14}, 
unveiling new two-dimensional (2D) materials as potential candidates
to be used in spin devices~\cite{Pesin12,Gao16,Nie17,Ashton17,Garcia18}. Specifically for graphene, half-metallicity in zig-zag nanoribbons \cite{Son06,Kan08}, 
defect-induced magnetism \cite{Yazyev07,Bundesman13} and spin transport at room temperature \cite{Tombros07, Han10}
have been predicted and experimentally demonstrated. 

{In addition to graphene, other column-IV 2D materials \cite{Molle17}, like silicene \cite{Lay12}, germanene \cite{Davila14} and lastly stanene \cite{Zhu16} have been recently pointed out to have an stable anti-ferromagnetic ground state when cut in thin zig-zag nanoribbons \cite{Wang16,Xiong16}.} A very promising property for spintronics in graphene, germanene and silicene is the electric-field-controlled half-metallicity,
i.e., the possibility to tune the nanoribbon bandgap from semiconducting to zero with a transversal in-plane electric field, that unfortunately is too large (from $10$~MV/cm to $25$~MV/cm) to be used in realistic applications and with practical dielectrics  \cite{Son06,Wang13}.

In this letter, we explore the appearance of half-metallicity in stanene 
nanoribbons with first principle calculations, and we propose a device based on interband 
tunneling able to exploit this effect to generate highly spin-polarized currents with 
small and realistic electric fields. We have opted for stanene amongst group-IV candidates because it
provides a reasonable compromise between the bandgap width (required to avoid high leakages currents) and 
the transversal electric field necessary to achieve half-metalicity  (that must prevent from the dielectric breakdown).
{In particular, the high sensitivity of  interband tunneling to small modulations of the bandgap and its reduced dependence on temperature, together with its robustness against the presence of defects \cite{Nanoscale17}, allows us to take advantage of the electric-field-controlled half-metallicity in a tunnel-field-effect transistor (TFET) based on a stanene nanoribbon.}

\begin{figure}[b]
	\includegraphics[width=0.6\columnwidth]{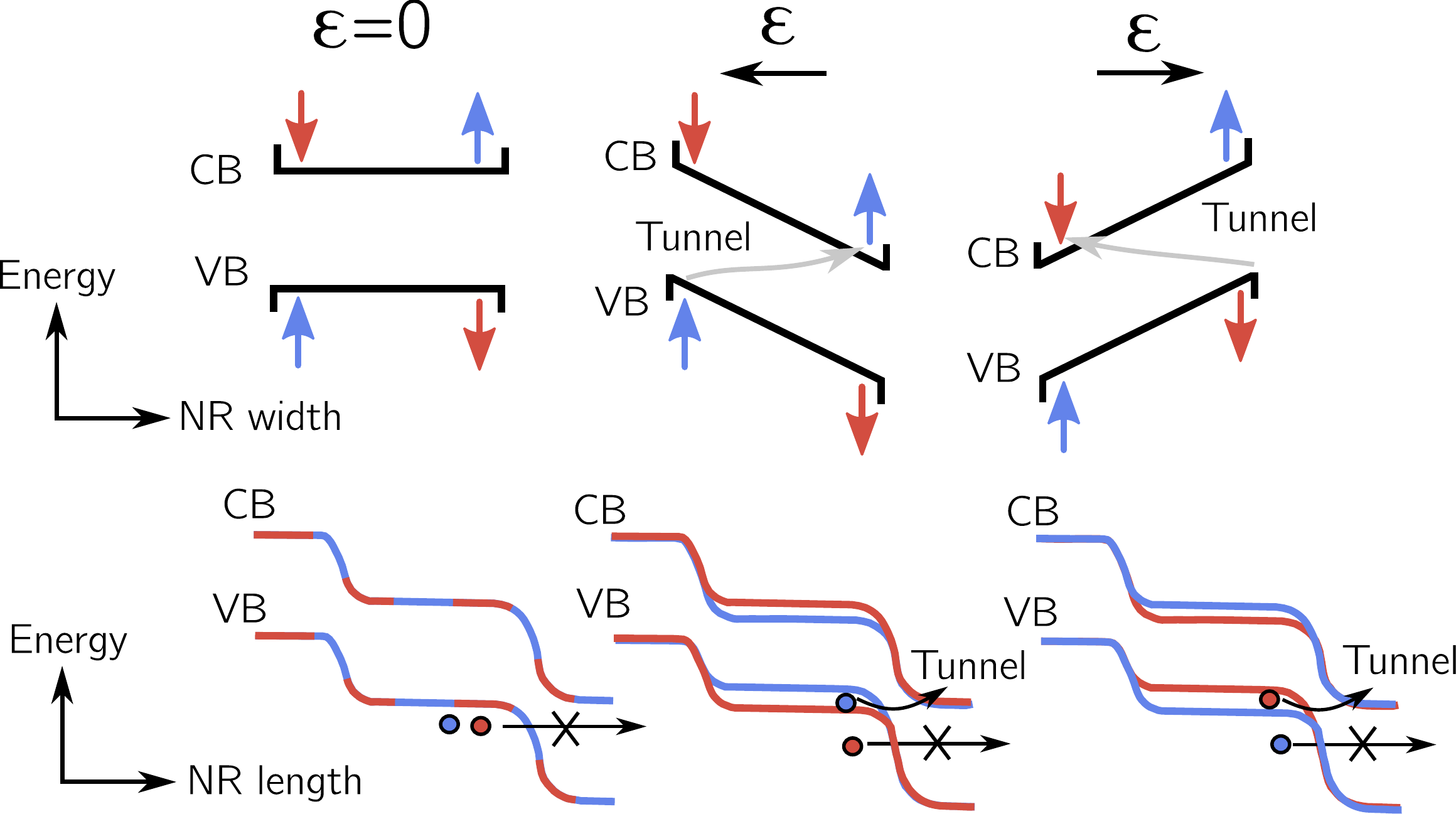}
	\caption{{Schematic of the device operation principle.} Schematic of the stanene nanoribbon (NR) conduction band (CB) and valence band (VB) along the nanoribbon width (top panel) and length (bottom panel) for spin-up (blue) and spin-down carriers (bottom), without/with an applied transversal electric field. }
	\label{fig:Fig00}
\end{figure}

A schematic illustration of device operation is shown in Fig. \ref{fig:Fig00}. {In a TFET, the source and drain regions are p- and n-doped respectively, pinning the chemical potential at the source/drain close to the valence/conduction band. At equilibrium, i.e, when no external voltages are applied, the chemical potential along the device is constant. When a positive drain-to-source voltage is considered, the chemical potential at the drain is lowered with respect to the source and an energy window for carrier tunneling is open. Then, when no transversal electric field is applied (left) the bandgap at any point along the channel is the same for both kinds of spins. If the source and drain regions are properly doped, the window for electron tunneling will face a thick barrier and the current will be small. An in-plane electric field applied in the transversal direction alters the bandgap for spin up and spin down carriers in the channel. As a consequence, the tunneling probability, from the valence band states in one edge to the conduction band states of the opposite edge, increases. The bandgap modulation and therefore the selective spin tunneling are due to the AFM configuration which is predicted to be preserved only for very thin nanoribbons ($\sim$1 nm).} The implementation of the band-gap modulation in a TFET  permits to obtain large spin-controlled currents with small modulations of the bandgap, or equivalently small electric fields, by properly aligning the undoped channel with the doped drain-end region (bottom panel).

\section*{Methods}
In order to demonstrate the operation of the proposed device concept we have adopted a multi-scale simulation approach, 
with ab-initio calculations of the stanene band-structure combined with self-consistent simulations of 
quantum transport and electrostatics. 

Density functional theory (DFT) as implemented in the Quantum Espresso suite~\cite{QE}, has been used to determine the electronic band-structure of a zig-zag stanene nanoribbon passivated at the edges with H atoms (Fig. \ref{fig:Fig0}).
For the DFT calculations, a space of $32$ \AA{} and $20$ \AA{} 
of vacuum in the direction of the nanoribbon width and the direction orthogonal to the nanoribbon plane, respectively, 
are assumed to minimize the interaction between periodic repetitions of the cell. We have performed a structural optimization 
within the Broyden-Fletcher-Goldfarb-Shanno algorithm until forces were smaller than $5\times 10^{-3}$eV$/$\AA{} 
with a convergence threshold for energy of $10^{-6}$ eV. A Perdew-Burke-Ernzerhof exchange-correlation functional 
is used \cite{PBE} within fully-relativistic norm-conserving pseudopotentials. 
Spin-polarized calculation, within the local spin density approximation (LSDA) and magnetization along the $\hat{z}$ axis is considered. 
The pseudopotentials are obtained from the SSSP Library, for both Sn and H \cite{SSSP}. 
Energy cuttoffs for charge density and wavefunction expansions are set to $360$ Ry and $30$ Ry, respectively. 
Integration in the Brillouin zone was accomplished in a $1\times10\times1$ $\Gamma$-centered grid. 

The device simulations are based on the self-consistent solution of the three-dimensional Poisson equation, 
together with the open-boundary Schr\"odinger equation, within the Non-equilibrium Green Functions 
(NEGF)~\cite{Datta00} formalism, formulated within a tight-binding (TB) scheme~\cite{TED18,VIDESwww}. To this purpose, we have projected the plane-wave DFT basis  set into a Maximally Localized Wannier Functions (MLWF) basis set, exploiting Wannier90 code~\cite{Wannier}, 
resulting in Hamiltonians of $96$ nearest-neighbors for both spin-up and spin-down states. Details of the adopted multi-scale approach can be found in ~\cite{Pizzi16,TED18}.

\section*{Results}

\begin{figure}[t]
 	\includegraphics[width=0.6\columnwidth]{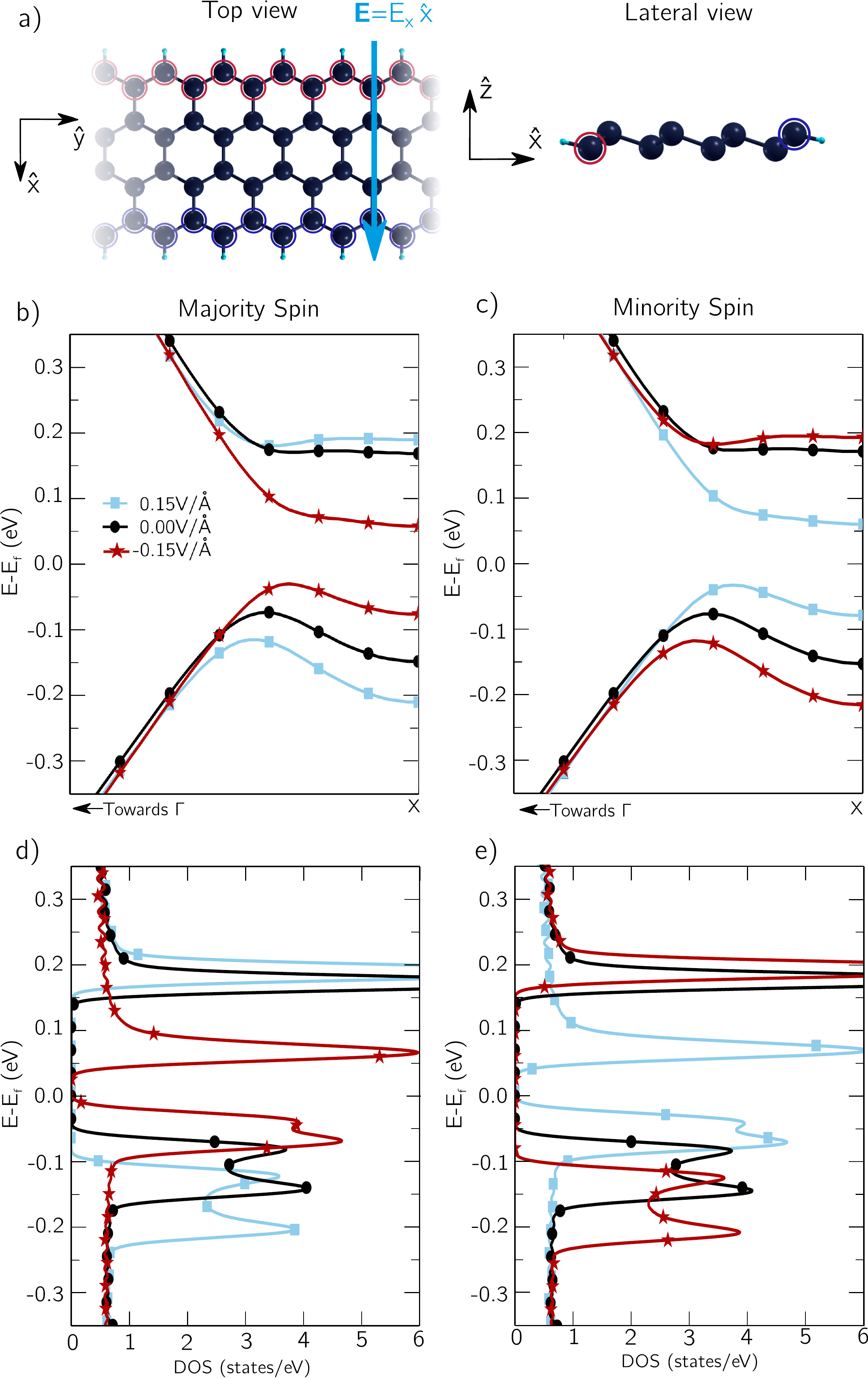}
 	\caption{{Crystal- and band-structure of the stanene nanoribbon.} a) Top and lateral view of the stanene zig-zag nanoribbon ($1.26$ nm wide) passivated with H atoms. The electric field, $\mathbf{E}=E_{\rm x}\hat{x}$ is applied along the nanoribbon width. b),c) Electronic band-structure and d),e) density  of states for spin-up (left) and spin-down (right) states for several electric fields.}
 	\label{fig:Fig0}
 \end{figure}

We have considered the stanene zig-zag nanoribbon shown in Fig. \ref{fig:Fig0}. 
As graphene, stanene is characterized by a honeycomb lattice (with unit vector of $4.7$~\AA{}), but with a buckled structure 
with two parallel planes of atoms separated by $0.82$~\AA{} \cite{Zhu16}. 
The considered nanoribbon width is $1.29$ nm: for such a thin nanoribbon, a large 
exchange interaction between edges is expected, preserving magnetism at room temperature \cite{Jung09,Yazyev10}.
{While spin orbit coupling  (SOC) is crucial for the appearance of topological spin-polarized edge states in topological insulators, in the anti-ferromagnetic configuration the spin polarization emerges due to the inter-edge exchange and SOC plays a minor role. In particular, for the studied system we have observed that the calculations including SOC have a negligible impact on the band-structure and the sensibility to the electric field is not modified.}

In the absence of electric field, the stanene bandstructure for spin up and down is degenerate 
(Fig. \ref{fig:Fig0}b and c, black circles), with an energy bandgap of $\sim$0.25~eV. 
The states close to the conduction band minimum (CBM) and the valence band
maximum (VBM) are localized at the edges of the nanoribbon, and according to the anti-ferromagnetic configuration, 
present opposite spin \cite{Xiong16,Nanoscale17}.
By applying a positive transversal electric field, $\mathbf{E}=E_{\rm x}\hat{x}$ (Fig. \ref{fig:Fig0}a), the energy 
bandgap increases for spin up bands and decreases for spin down bands: opposite behavior occurs for negative  $\mathbf{E}$.
For example, for $E_{\rm x}=0.15$~V/\AA{}, the spin-down bandgap reduces to $0.087$~eV, 
and for spin-up increases to  $0.296$~eV (Fig \ref{fig:Fig0}b and c).

As can be seen in Fig.~\ref{fig:Fig0}d and e, the small curvature of the stanene bandstructure 
around the CBM and the VBM leads to corresponding peaks in the density of states (DOS). 
Coherently with Fig.~\ref{fig:Fig0}b and c the spin-up 
and spin-down DOS have opposite dependence on $E_{\rm x}$. Eventually, for $E_{\rm x}=\pm 0.3$V/\AA{} 
the material becomes a half metal (zero bandgap). A similar magnitude of the electric field 
is necessary to achieve half-metallicity
in graphene, silicene, and germanene \cite{Son06,Wang13}. These values are, however, larger than the breakdown fields of practical dielectrics, 
and a device design sensitive to a small modulation of the bandgap 
is mandatory to take practical advantage of this effect in spintronics.

\begin{figure} [t]
 	\includegraphics[width=0.6\columnwidth]{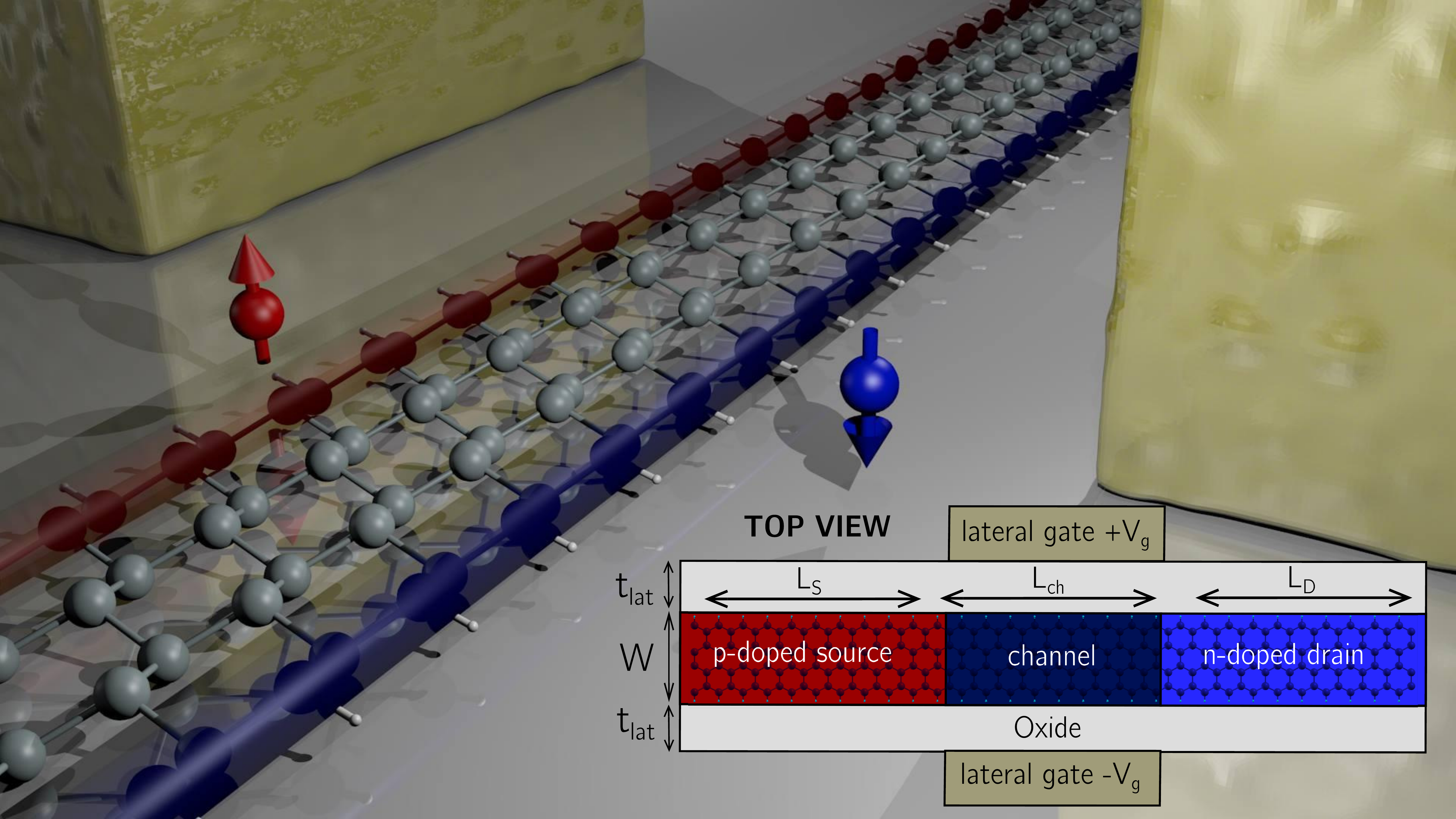}
 	\caption{{Schematic of the proposed device.} The stanene zig-zag NR is embedded in top, bottom, and lateral SiO$_2$, with thicknesses $t_{\text{ox}}=1$ nm and $t_{\text{lat}}=1$ nm. Two lateral metallic contacts control the spin current flow. Inset: top view of the device. $L_{\rm ch}=16.9$~nm, $L_{\rm S/D}=5.6$~nm and a doping molar fraction $1.2\times 10^{-2}$ has been considered for the $p-/n-$type source/drain.}
 	\label{fig:Fig1}
 \end{figure}

To this purpose, we propose a TFET structure where the stanene nanoribbon
is embedded in SiO$_2$, and sandwiched between two lateral metallic contacts, placed $1$~nm 
far from each nanoribbon edge (Fig.~\ref{fig:Fig1}). The current flow is enabled by  
interband tunneling between the undoped-channel valence band (VB) and the doped-drain 
conduction band (CB), but - differently from a TFET - the switching current modulation 
is due to the modulation of the channel bandgap.
The interband tunneling current is extremely sensitive to bandgap variations, so that with small 
electric fields it is possible to tune the alignment between the VB and the CB and consequently 
achieve large modulation of the spin-polarized current. 

The considered nanoribbon has a length of $28.2$~nm. At its ends, we have assumed $5.64$~nm-long 
$p$-/$n$-doped regions, with acceptor/donor molar fractions 
equal to $1.2\times 10^{-2}$. The $16.9$~nm-long central region is the device channel.  
The two lateral gates are used to create a transversal electric field that,  
depending on its sign, allows to open/close the channel bandgap for each spin 
component.

In order to accurately study the performance of the proposed device, we have performed transport simulations 
solving self-consistently the three-dimensional Poisson equation, 
together with the open-boundary Schr\"odinger equation (see Methods). Spin-up and spin-down  polarized currents ($I_{\rm ds}^{\rm \uparrow/ \downarrow}$)  in the ballistic regime at room temperature 
($T=300$ K) have been calculated exploiting the Landauer's formalism \cite{Landauer88,Buttiker90}.

\begin{figure} [b]
 	\centering
 	\includegraphics[width=1\textwidth]{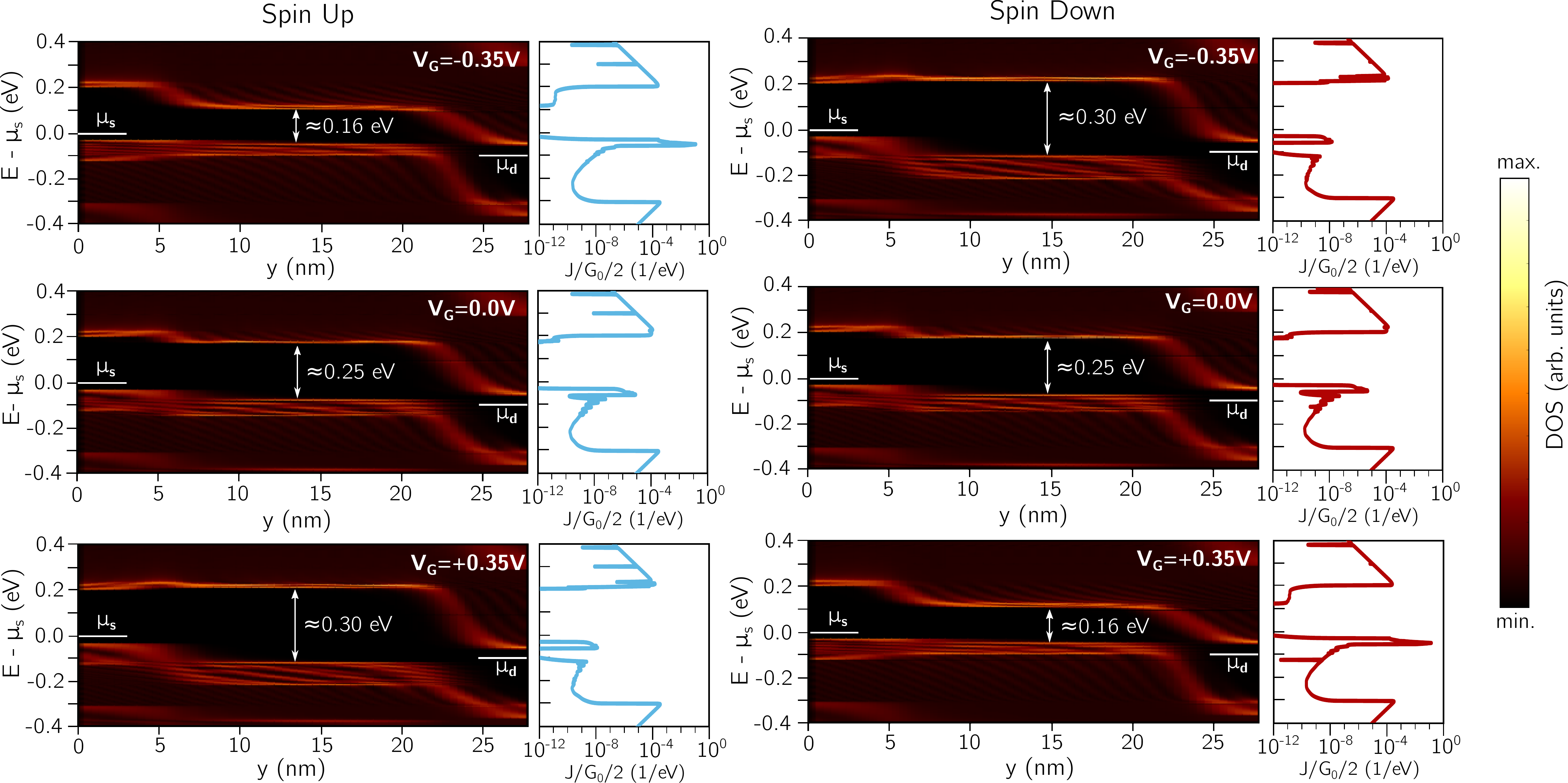}
 	\caption{{Density of states and current-density spectrum for several bias voltages.} Density of states (DOS) as a function of energy and longitudinal position along the nanoribbon for the spin-up (left) and spin-down (right) carriers for several applied voltages ($V_{\text{G}}$), for a drain-to-source voltage of $V_{\text{ds}}=0.1$~V. Aside to each DOS colormap: current-density spectrum in semilogarithmic scale normalized to half the conductance quantum.}
 	\label{fig:Fig2}
 \end{figure}

In order to illustrate the working principle of the proposed device, we have depicted the density of states (DOS) as a function of the energy (refereed to the Fermi level at the source lead), and the position along the nanoribbon length ($y$), for both spin-up and spin-down carriers (Fig. \ref{fig:Fig2}). 
On the right of each DOS plot we have included the spectrum of the current density for the corresponding spin, normalized to half the conductance quantum ($G_0/2=q^2/h$) in semi-logarithmic scale. A drain-to-source voltage $V_{\rm ds}=0.1$~V and several values of $V_{\rm G}$ are considered. The Fermi level at the source ($\mu_{\rm s}$) and drain ($\mu_{\rm d}$) leads are depicted together with the gap value at the channel. 
From the DOS colormap, one can easily recognize the CB and the VB, that correspond to two high-DOS stripes. 
The energy gap is also identified as a zero-DOS region between the CB and the VB. 
As can be seen, in the channel region ($5.6$~nm$<y<22.5$~nm) the bandgap width is modulated by $V_{\rm G}$, while it remains invariant at the 
source and drain regions where the influence of the gates is negligible.  
The device behavior is thus explained as follows: when $V_{\rm G}=0$~V (Fig. \ref{fig:Fig2} center panel) 
spin-down and spin-up carriers face equal barriers for inter-band tunneling and the current-density spectrum {(normalized to $G_0/2$)} does not show significant differences,
and therefore the total current has no spin polarization.  
On the contrary, when $V_{\rm G}\neq0$~V spin-up and spin-down carriers experience different channel interband tunneling bandgap 
and consequently the current has a spin polarization.

In particular, for $V_{\text{G}}=-0.35$~V (Fig. \ref{fig:Fig2} top panel), the bandgap for the spin-down states widens ($\sim0.30$~eV) 
whilst it closes ($\sim0.16$~eV) for spin-up states. As a consequence, when the channel bandgap is increased the inter-band 
(from the channel VB to the drain CB) tunneling probability is reduced, resulting in a 
 decrease of the spin-down current, until only the thermionic components are visible.  
 On the other hand, as the channel bandgap is reduced for spin-up states, the carriers from the 
 $p$-doped source VB fill the channel VB, and tunnel through the channel-VB-to-drain-CB thin barrier, notably
 increasing the tunneling current for this spin component, as shown by the current-density spectrum, 
 and resulting in a highly spin-polarized total current.  Symmetrically opposite behavior is observed for $V_{\text{G}}=0.35$ V (Fig. \ref{fig:Fig2} bottom panel).

\begin{figure} [h]
	\includegraphics[width=0.6\columnwidth]{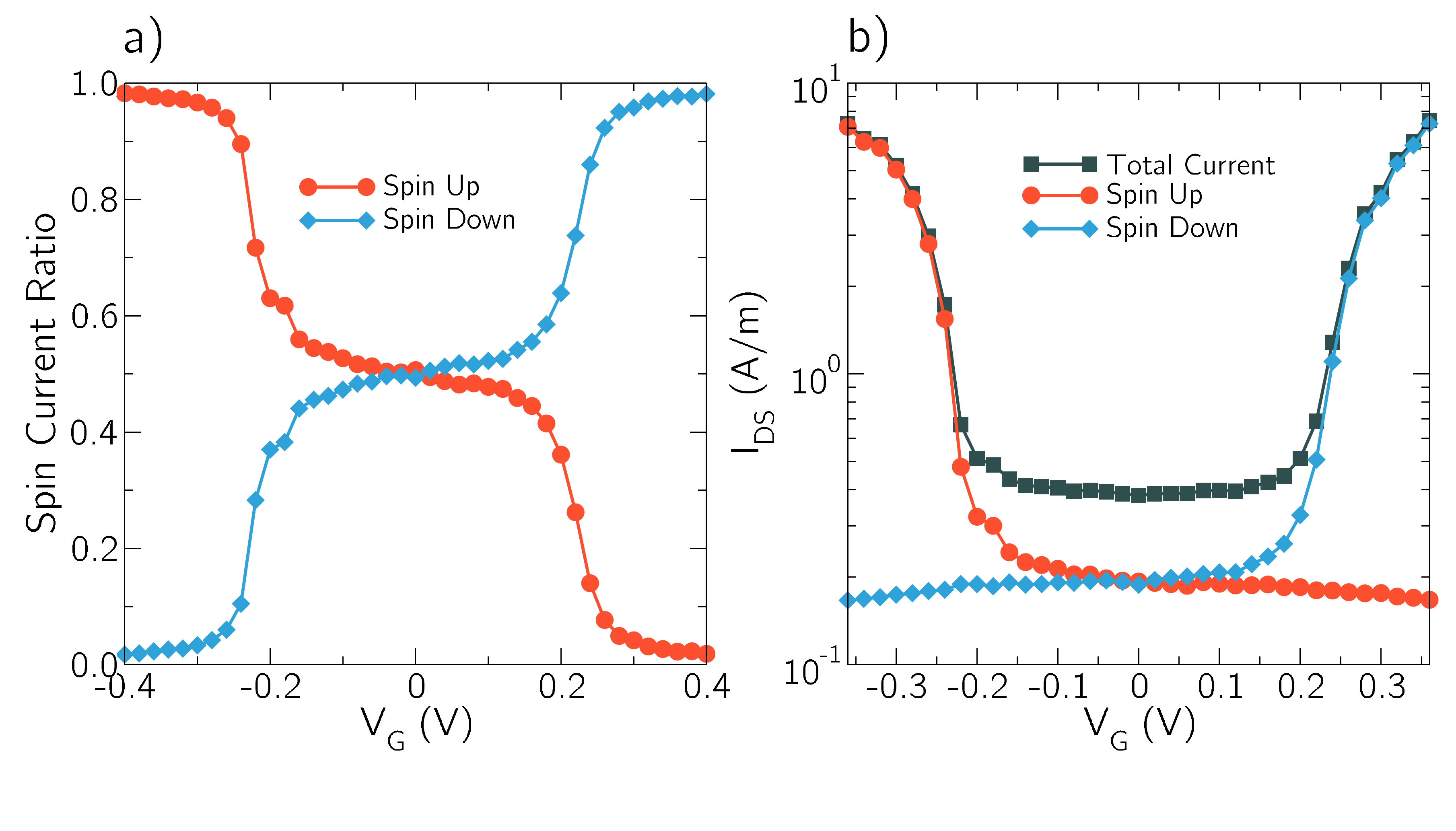}
	\caption{{Spin currents.} a) Spin-to-total current ratio and b) Total current and spin components in semilogarithmic scale, as a function of the voltage applied at the lateral gates ($V_{\text{G}}$), for a drain-to-source voltage of $V_{\text{ds}}=0.1$~V.}  
	\label{fig:Fig3}
\end{figure}

The spin-to-total current ratio as well as the total current and the spin-up and spin-down components are shown as a function of the applied gate voltage in Fig. \ref{fig:Fig3}, for $V_{\rm ds}=0.1$~V. The total current (grey squares) shows a symmetric behavior 
with a minimum at $V_{\rm G}=0$ V, when current has no spin polarization, as spin-up (red circles) and spin-down (blue triangles) 
currents provide equal contribution. If a non-zero $V_{\text{G}}$ is applied, the symmetry is broken, and 
the total current becomes slightly spin-polarized with spin-up for $V_{\text{G}}>0$ and spin-down for $V_{\text{G}}<0$. 
The spin polarization of the current increases with $V_{\text{G}}$ and achieves $90\%$ 
for $V_{\rm G}=\pm 0.2$V, and $98\%$ for $V_{\text{G}}= \pm 0.3$ V.

\section*{Discussion}

We have theoretically demonstrated the electrical behavior of a device 
based on 2-D stanene, able to provide tunable spin-polarized current up to a polarizion of 98\% 
with a small applied voltage on the controlling electrode.
The proposed device is based on the modulation of the bandgap of narrow stanene 
nanoribbons and the edge-localized nature of the conduction and the valence band states.
Taking advantage of this property, the device controls the interband tunneling currents 
through electric-field modulation of the energy bandgap, achieving a spin polarization as 
high as $98\%$ with electric field of 0.3~V/nm. We think that the proposed device might 
be useful to explore new concepts of spin injectors or filters, 
that are fundamental building blocks of spintronics.

\section*{Acknowledgements} Authors gratefully acknowledge the support from the Graphene Flagship Core 2 Contract No. 785219.

\end{document}